\documentclass[pre, twocolumn, amsmath, amssymb, superscriptaddress]{revtex4}

\usepackage{gensymb}

\usepackage{graphicx}
\usepackage{dcolumn}
\usepackage{bm}
\usepackage{braket}
\usepackage{xcolor}
\usepackage{verbatim}
\usepackage{amsmath}

\newcommand{\beq}{\begin{equation}}
\newcommand{\eeq}{\end{equation}}

\begin{document}

\title{ A mother-machine microfluidic device for non-adherent mammalian cells reveals the population growth strategies}



\author{Simone Scalise}
\affiliation{Center for Life Nano \& Neuro Science, Istituto Italiano di Tecnologia, Viale Regina Elena 291,  00161, Rome, Italy}

\affiliation{Department of Physics, Sapienza University, Piazzale Aldo Moro 5, 00185, Rome, Italy}

\author{Giovanna Peruzzi}
\affiliation{Center for Life Nano \& Neuro Science, Istituto Italiano di Tecnologia, Viale Regina Elena 291,  00161, Rome, Italy}


\author{Davide Caprini}
\affiliation{Center for Life Nano \& Neuro Science, Istituto Italiano di Tecnologia, Viale Regina Elena 291,  00161, Rome, Italy}

\author{Giorgio Gosti}
\affiliation{Institute of Heritage Science, National Research Council (CNR-ISPC), Via Salaria KM 29300, 00015 Monterotondo, Italy}
\affiliation{Center for Life Nano \& Neuro Science, Istituto Italiano di Tecnologia, Viale Regina Elena 291,  00161, Rome, Italy}

\author{Giancarlo Ruocco}
\affiliation{Center for Life Nano \& Neuro Science, Istituto Italiano di Tecnologia, Viale Regina Elena 291,  00161, Rome, Italy}

\affiliation{Department of Physics, Sapienza University, Piazzale Aldo Moro 5, 00185, Rome, Italy}

\author{Mattia Miotto  \footnote{\label{corr} For correspondence write to: mattia.miotto@roma1.infn.it}}
\affiliation{Center for Life Nano \& Neuro Science, Istituto Italiano di Tecnologia, Viale Regina Elena 291,  00161, Rome, Italy}

\begin{abstract}
We develop a mother machine-like microfluidic device specifically designed to track the proliferation of T-cells via live-cell microscopy.
Although numerous microfluidic setups have been developed to study cell proliferation at the single-cell level, most of them are optimized for use on adherent cells.
Here, we present a device to track the proliferation of suspension cells, featuring an array of microchannels that trap cells, easing their monitoring while allowing for controlled growth conditions.
Each microchannel, whose geometry has been optimized through computational fluid dynamics simulations, allows a single cell to enter and proliferate while maintaining a continuous flow of nutrients, ensuring long-term monitoring over multiple generations. 

We show the advantages of this system in  characterizing 
the proliferation of human leukemia T-cells. In particular, we 
follow the growth and division over multiple generations, finding that 
 cells exhibit a slightly asymmetric volume division where deviations in the size are compensated by a size-like division strategy.
Overall, our device design can be easily adapted and used to study different cell types and sizes while maintaining the same high trapping efficiency.
\end{abstract}

\maketitle

\section{Introduction}

Understanding the mechanisms that regulate cell growth and division is one of the central questions in cell biology \cite{jun2015cell}. The strategies cells adopt to decide when and how to divide can vary significantly depending on their physiological state and cell type \cite{kussell2005phenotypic, DeMartino2019}. In cancer cells, for instance, cell cycle progression is often deregulated, resulting in highly heterogeneous proliferative behaviors \cite{buss2024contribution}.
Traditional approaches to study these dynamics, such as flow cytometry, typically provide population-level measurements, which preclude the reconstruction of lineage histories: while flow cytometry can be used to infer the growth strategies adopted by cells~\cite{miotto2023determining, scalise2024}, it is intrinsically limited in its ability to capture dynamic behaviors at the single-cell level \cite{longo2006dynamics, perie2016retracing}.
This has led to a growing demand for methods that can monitor dynamic processes over time, particularly in biologically heterogeneous systems such as cancer cell populations. 
In this context, time-lapse microscopy coupled with microfluidic devices has become an essential tool, offering a powerful method to investigate phenotypic variability and dynamic processes at the single-cell level \cite{halldorsson2015advantages}. 
In this approach, cells are typically cultured under a microscope in a controlled environment and imaged at regular intervals for long periods, allowing for the tracking of individual growth trajectories and division events. While this is relatively straightforward for adherent cells, which remain immobilized on the imaging plane, suspension cells pose a significant challenge due to their inherent motility \cite{ivanusic2017easy}. Even small perturbations, such as medium perfusion or thermal drift, can disrupt cell positioning, compromising long-term imaging and lineage reconstruction.
The advent of microfluidic devices allowed to overcome many of that limitations. In fact, such systems permit precise control over cell positioning and environmental conditions, including nutrient delivery and waste removal, with minimal perturbation. In recent years, a variety of microfluidic platforms have been developed for different applications, including morphological characterization \cite{chen2011classification, hong2012electrical, lam2017adaptation}, viability assays \cite{komen2008viability,wheeler2003microfluidic}, and studies of proliferation, cell-cell interaction \cite{ramadan2013nutrichip,zheng2012quantitative}, or response to stimuli \cite{chung2011imaging, el2005cell}. Depending on the architecture, these systems may focus on single-cell resolution or the behavior of small populations under tightly controlled conditions.

\begin{figure*}
    \centering
    \includegraphics[width=\textwidth]{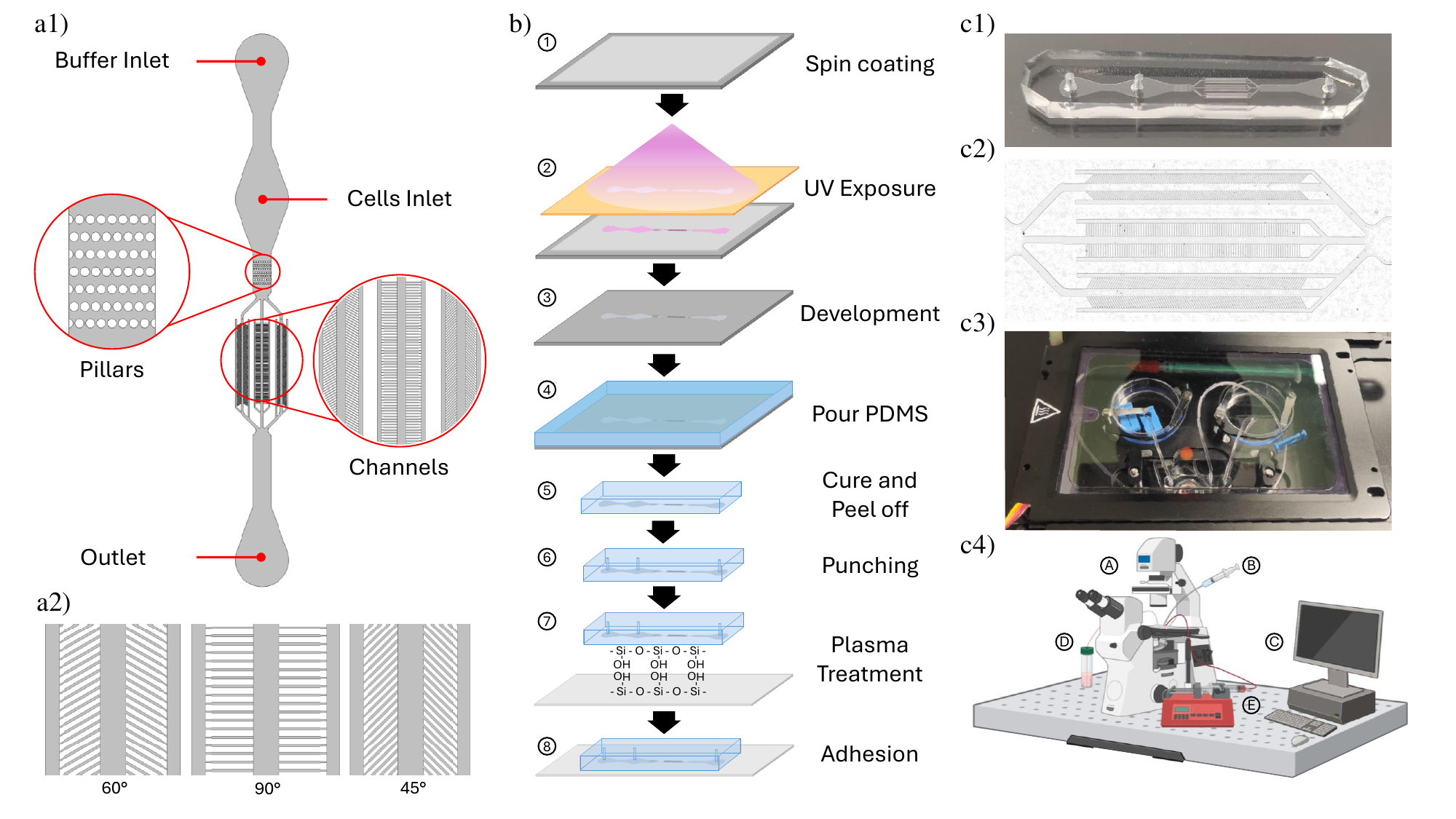}
    \caption{\textbf{Chip design and fabrication.} 
    \textbf{a1)} Schematic representation of the microfluidic chip specifying the main components. 
    \textbf{a2)} A zoom on the three different channel orientations within the device. From left to right, we have channels with 60, 90 and 45 degrees slope to the main flow channel, respectively.  
    \textbf{b)} Schematization of microfabrication steps (see Materials and Methods - Microfabrication for more details).
    \textbf{c1)} A picture of the fabricated microfluidic device. The PDMS layer is irreversibly bonded to the glass slide.
    \textbf{c2)} Picture of the trapping channels obtained by a mosaic of the 10x frames made at the microscope.
    \textbf{c3)} Photo of the chip inside the top-stage incubator with the inlet and outlet tubes connected.
    \textbf{c4)} Schematic representation of the experimental setup (see Materials and Methods - Experimental setup for more details).
    }
    \label{fig:1}
\end{figure*}

Current microfluidic designs can be broadly categorized based on their cell confinement strategy. Devices based on hydrodynamic or geometric trapping offer high-resolution imaging of individual cells but are typically incompatible with studies of proliferation or long-term lineage tracking \cite{chai2021microfluidic,jin2015microfluidic,ahmad2016microfluidic, wheeler2003microfluidic, di2006dynamic}. On the other hand, barrier-based systems employ physical structures to spatially constrain cell populations, facilitating the analysis of collective dynamics, but often at the cost of reduced perfusion and poor compatibility with suspension cultures \cite{karakas2017microfluidic, lin2015microfluidic}. A hybrid approach is represented by chamber-based systems, which provide enough space for both initial trapping and subsequent proliferation while allowing for medium perfusion. Despite their success in bacterial \cite{bakshi2021tracking,sun2011high, hornung2018quantitative}, yeast \cite{jo2015high,liu2015yeast,crane2014microfluidic}, or adherent mammalian models \cite{kolnik2012vacuum,wang2009self}, few of these devices have been adapted for non-adherent mammalian cells. Existing solutions for suspension cultures often complicate long-term cell tracking due to less structured spatial confinement \cite{cambier2015design,keenan2010new}. 
In this work, we introduce a microfluidic device specifically designed for the long-term cultivation and tracking of suspension cancer cells, with a focus on Jurkat T lymphocytes. Inspired by the  ``mother machine" design originally developed for bacterial lineage analysis \cite{wang2010robust,si2019mechanistic,thiermann2024tools, taheri2015single}, our device consists of a dense array of narrow microchannels, each ideally capturing a single cell. Over time, cells proliferate along the length of the channel, enabling the reconstruction of full lineage trees across multiple generations. The geometry of the device ensures both physical confinement and sustained medium exchange, making it suitable for dynamic single-cell studies under near-physiological conditions. Our direct measurements of cell partitioning noise, division strategy, and proliferation capacity are in quantitative agreement with previously reported findings obtained with indirect methods. The obtained results demonstrate that this platform allows for the reliable tracking of proliferating cells over extended periods and confirms and indicate the device as a promising tool for studying the dynamics of T-cells.

\begin{figure}
    \centering
    \includegraphics[width=\columnwidth]{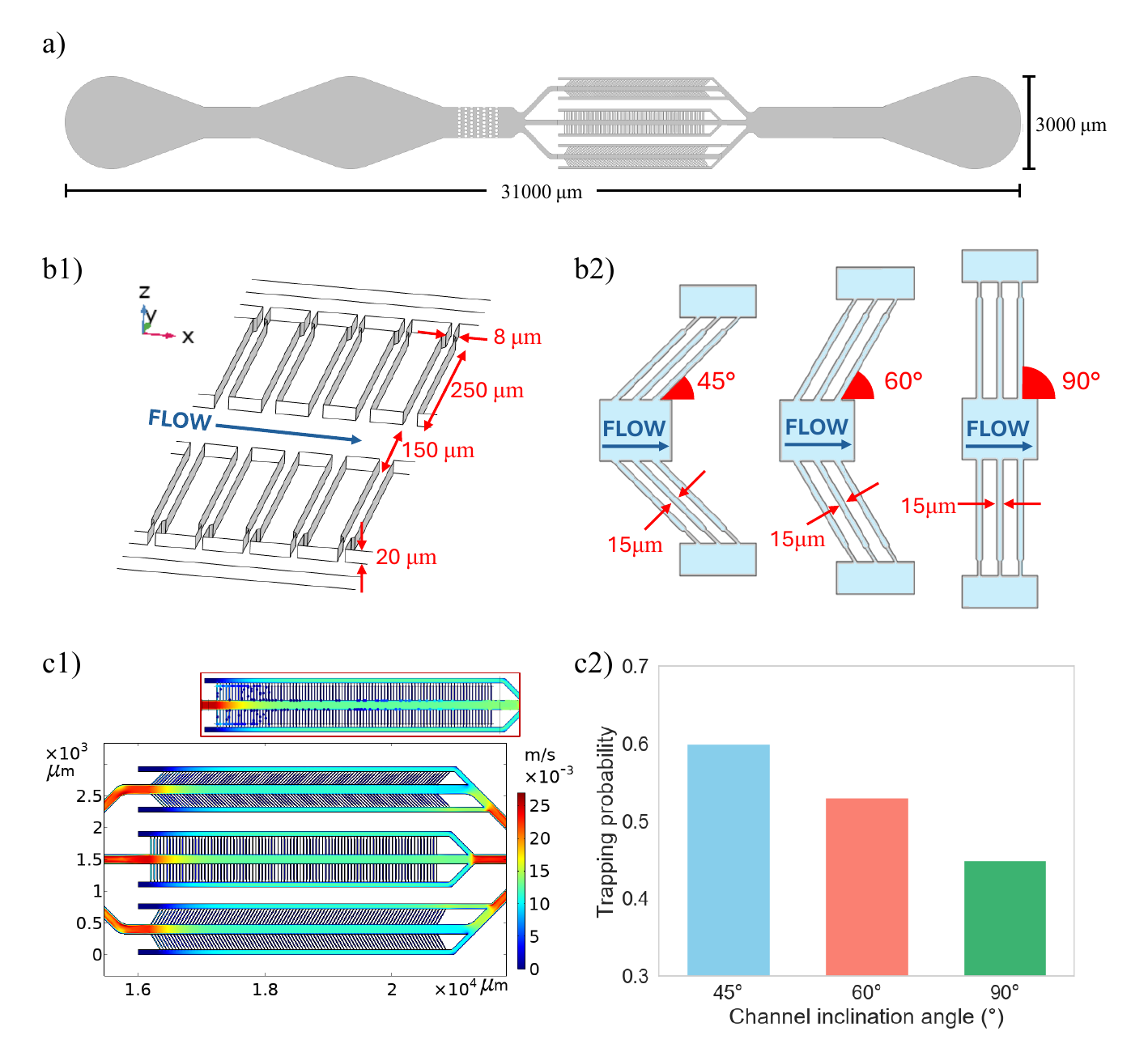}
    \caption{\textbf{Component effect and dynamics.} 
    \textbf{a)} General layout of the microfluidic device with dimensions shown on the side.  
    \textbf{b1)} Diagram of trapping channels oriented 90 degrees from the main channel with the dimensions of the various sections in $\mu$m. 
    \textbf{b2)} Schematic representation of the inclination of the side channels with respect to the main channel. All side channels share the same width and length.  
    \textbf{c1)} Computational simulation results for the velocity field of laminar flow (without particles or cells) through the trapping zone. The section was taken in the horizontal median plane (at half of the channel height). Inset showing a frame from a trapping simulation. Cells are represented as blue dots, while the velocity field is the same as in the main panel.  
    \textbf{c2)} Histogram of trapping probabilities as a function of channel inclination, obtained from computational simulations.  
    }
    \label{fig:2}
\end{figure}

\begin{figure*}
    \centering
    \includegraphics[width=0.8\textwidth]{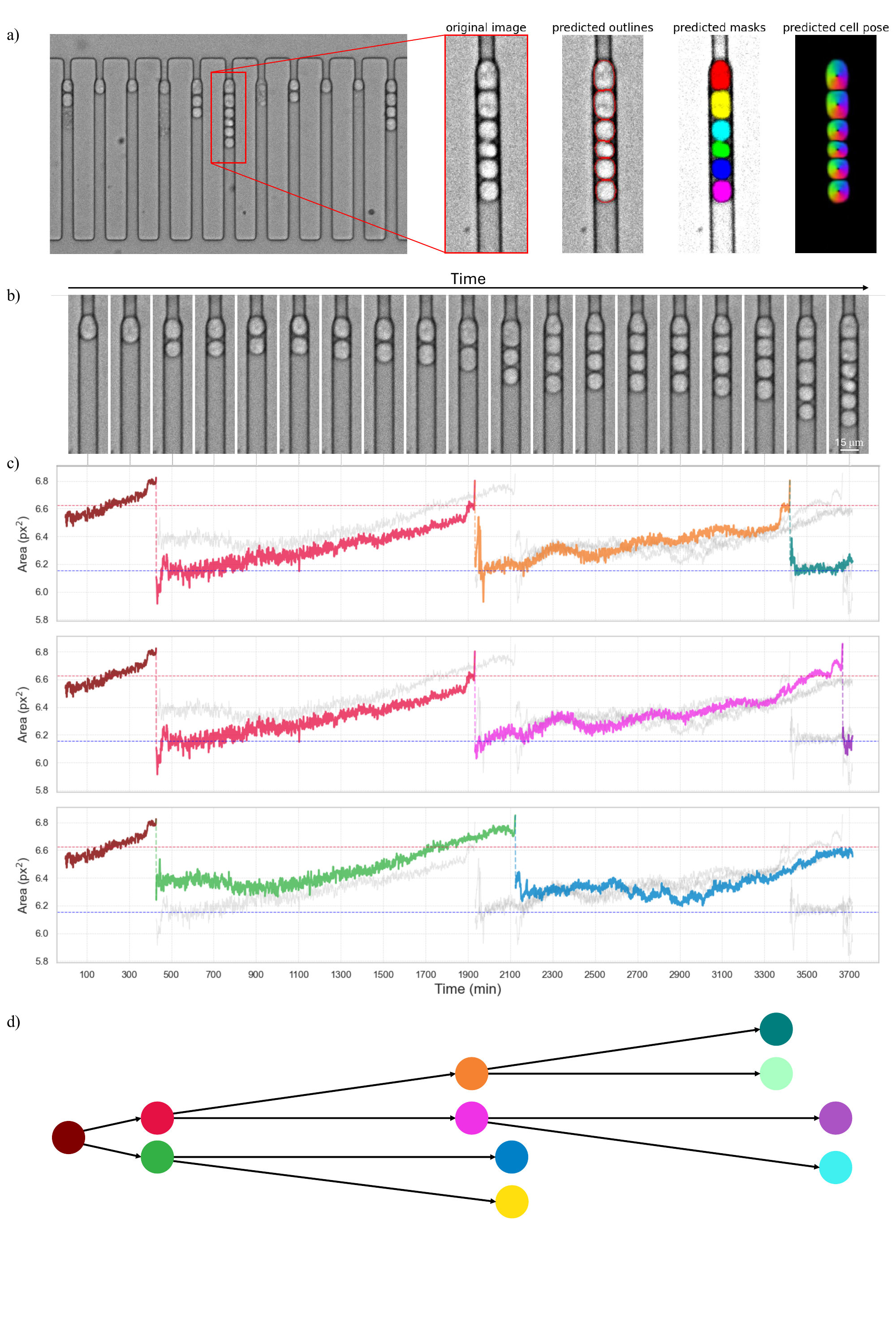}
    \caption{\textbf{Segmentation and lineage tracking of single cells over time.} 
    \textbf{a)} Representation of the segmentation process applied to the channels. Once the channel containing cell divisions is identified, it is analyzed with Cellpose. The algorithm first detects cell boundaries and generates the corresponding masks. Subsequently, a vector flow field is computed, representing the spatial gradients that point from each pixel toward the center of the nearest cell. By integrating these fields, Cellpose reconstructs accurate cell contours.  
    \textbf{b)} Frames of the same channel at different time intervals in which cell proliferation from a single cell can be observed.
    \textbf{c)} Plot of cell growth and division over time from the images shown above. Starting from a single cell (mother), we observe growth followed by division into two daughter cells, which subsequently divide into two granddaughter cells at different times. In two cases, further divisions occur, leading to great-granddaughter cells. The three panels display three of the six distinct lineages identified.  
    \textbf{d)} Tree schematization of cell genealogy of the previous plot. The same colors are associated with the same cell, while the arrows' length corresponds to the division time. 
    }
    \label{fig:3}
\end{figure*}

\begin{figure*}
    \centering
    \includegraphics[width=\textwidth]{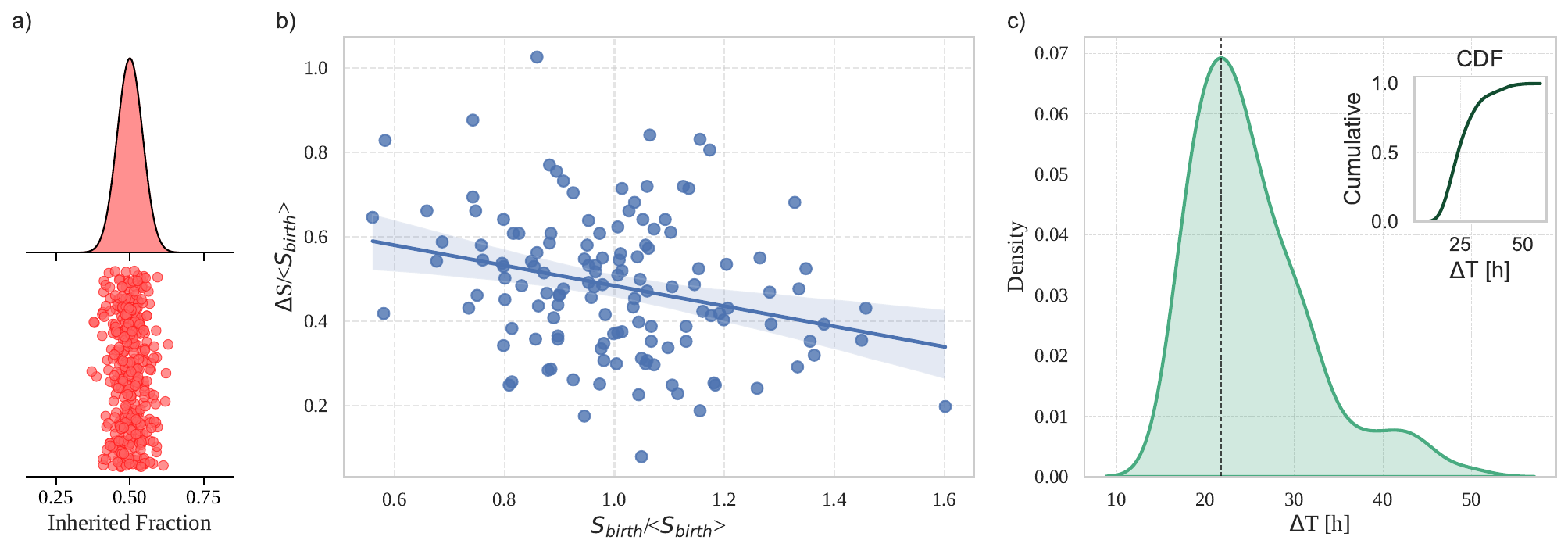}
    \caption{\textbf{Growth regulation and division dynamics.} 
    \textbf{a)} Scatter plot (bottom) and relative fit distribution (top) of inherited volume fraction (defined as the ratio of the area of a daughter cell to the sum of the areas of both daughters) for all observed cell divisions. The distribution is a Gaussian with $\mu = 0.5$ and $\sigma = 0.05$.
    \textbf{b)} Variation of size as a function of birth size for all observed cell cycles. Values are normalized to the average birth size. From the slope of the regression line, $m=-0.24$, it can be inferred that cells follow a sizer-like growth model.
    \textbf{c)} Distribution of division times. The black dashed line corresponds to the peak of the distribution at $\Delta h = 22$ h. The inset shows the cumulative distribution, from which it can be observed that only 10\% of cells exhibit a division time longer than 35 h.
    }
    \label{fig:4}
\end{figure*}

\begin{figure*}
    \centering
    \includegraphics[width=\textwidth]{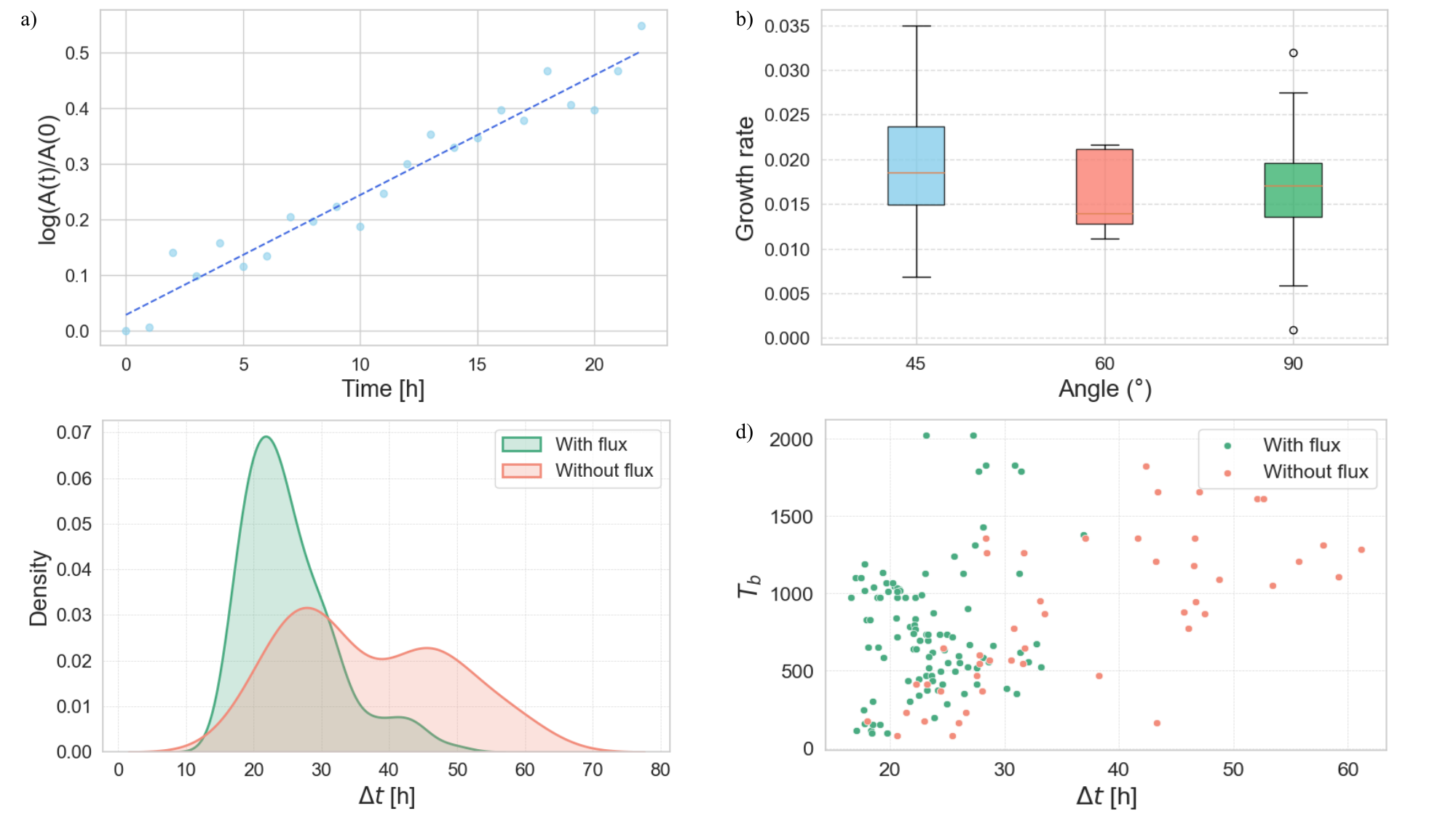}
    \caption{\textbf{Impact of channel geometry and nutrient supply on cell growth and division.} 
    \textbf{a)} Example of a single-cell growth curve in a channel. The cell was segmented every hour from birth to division. Values were normalized to the initial size at $t=0$ and plotted on a logarithmic scale to highlight exponential growth.
    \textbf{b)} Distribution of growth rates of cells as a function of the channel inclination. Among the three configurations, channels at $45^{\circ}$ exhibit, on average, the highest growth rates.
    \textbf{c)} Division-time distributions from experiments performed with or without medium flow during the time-lapse. Compared to cells under flow, those grown without flow display an overall right-shifted distribution, with longer division times, characterized by a first peak at $\Delta h = 28$ h and a secondary peak at $\Delta h = 46$ h.
    \textbf{d)} Scatter plot of birth time versus division time for proliferating cells under flow and no-flow conditions. In the absence of flow, cells born later in the experiment exhibit substantially longer division times compared to those born within the first 18–20 h. By contrast, in the presence of a constant nutrient supply, division times remain below 40 h regardless of birth time.
    }
    \label{fig:5}
\end{figure*}

\section{Results and Discussions}

\subsection{Design of the device}

The device was conceived to enable long-term monitoring of the proliferation of suspension cells. Achieving this required a design that could physically retain the cells for time-lapse imaging, while still allowing sufficient spatial freedom to support their growth. To this aim, we developed a microfluidic chip featuring a comb-like arrangement of microchannels, acting as cell traps. Unlike the closed-ended channels commonly used in existing devices, primarily for bacterial colony studies \cite{wang2010robust,bakshi2021tracking,yang2018analysis, norman2013memory}, we opted for channels open at the distal end and connected to a secondary drainage channel through a narrow constriction (see Figure~\ref{fig:1}).

This setup plays a key role, i.e. it generates flow lines that establish a unidirectional current through the microchannels. This flow serves two purposes. First, it drives cells into the channels while preventing their escape, eliminating the need for centrifugation-based loading, which can induce mechanical stress and reduce viability. Second, it ensures continuous medium exchange, providing nutrients to all cells within the channel. Without this flow, nutrient replenishment would rely solely on diffusion from the main channel, benefiting primarily the outermost cells while depriving those deeper in the channel.

To optimize cell loading, we adopted a dual-inlet configuration. As previously reported~\cite{Dualinlet, sun2016comparison}, single-inlet designs can lead to cell accumulation near the loading port, which poses several problems: cells can later enter the channels and be mistaken for division events, and excessive accumulation can obstruct normal medium flow. In our design, one inlet is dedicated solely to medium perfusion, followed downstream by a second inlet for cell loading. This arrangement avoids zero-flow zones near the cell inlet, reducing the likelihood of cell deposition in undesired locations.
Immediately after the loading-inlet, the flow passes through a region containing staggered pillar arrays. The spacing between pillars in each row is set to 20 $\mu m$, slightly larger than the average cell diameter, allowing at most two cells to pass simultaneously. This ensures that any remaining cell aggregates are retained in this region, preventing obstruction of the downstream trapping channels \cite{kaminaga2015uniform}. 

The device incorporates three distinct trapping channel geometries, differing in their orientation relative to the main flow channel: 
$45\degree$, $60\degree$, and $90\degree$. As discussed in detail in the following, these variations were designed to investigate how flow velocity affects cell entry and trapping efficiency. Differences in flow orientation also translate into variations in the mechanical stress experienced by the cells in each channel type.

\subsection{Optimization of the device}

The proposed device has been optimized for T-cells. In particular we have studied the Jurkat cells, an immortalized line of human T lymphocytes, characterized by a  spherical morphology and an average diameter of approximately 15 $\mu m$. This required adapting the dimensions of the device components to match the cell size. Since some design features described earlier, such as the trapping channels and the pillar arrays, can be easily scaled for cell types of different sizes, here they were tailored for Jurkat cells.

As shown in Figure~\ref{fig:2}b1 and ~\ref{fig:2}b2, all trapping channels, regardless of their inclination, were designed with a height of $\sim$ 20 $\mu m$ and a width of 15 $\mu m$, corresponding exactly to the cell diameter. This choice was based on preliminary tests with larger dimensions, which revealed that cells could reorient during division, aligning transversely to the channel axis (either stacked vertically or side-by-side). Such orientations complicated the identification of cell divisions and accurate size measurements during image segmentation. Conversely, dimensions smaller than the cell diameter hindered cell entry into the channels and caused excessive deformation, flattening the cells.

The channel length was set to allow tracking of multiple cell divisions, with sufficient space to accommodate up to the fifth generation of cells. Shorter channels would not allow enough generations to be monitored, limiting the ability to study correlations between successive cell cycles.

Considering cell size, the dimensions and spacing of the pillar arrays were also optimized to prevent the passage of cell aggregates while allowing single cells to flow freely. A spacing of 20 $\mu m$, slightly larger than the cell diameter, proved effective in blocking clusters and debris without adding unnecessary mechanical stress, which would occur with narrower gaps.

Preliminary tests also informed the optimization of the main channel length by increasing the number of trapping channels for each inclination. Figure~\ref{fig:2}c1 shows the results of fluid-dynamics simulations (see Methods for details), done to guide the optimization process. Increasing the length of the main channel (and thus the total number of trapping channels) reduces the local flow velocity, enabling cells to enter the channels via Brownian motion rather than solely through flow-driven capture. The inset in the same figure illustrates this principle: at the beginning of the main channel, particle entry is dominated by flow, whereas towards the end it is primarily driven by diffusion.

Moreover, making the device excessively long would be impractical, as it would significantly increase the area to be imaged in time-lapse experiments, leading to longer acquisition times per measurement and larger data volumes to store and process.

Finally, trapping efficiency was also investigated by incorporating three different channel inclinations ($45\degree$, $60\degree$, and $90\degree$) within the same device. Channel inclination primarily affects the local flow velocity inside the trapping channels. Computational simulations releasing showed the highest entry probability (about 60\%) for $45\degree$ channels, decreasing to about 45\% for perpendicular channels ($90\degree$) (see Fig~\ref{fig:2}c2). 

\subsection{Live-cell timelapse microscopy experiment}
Next, we turned to testing our device using Jurkat T-cells, an immortalized  line of human T-cells lymphocyte from a patient affected by acute T-cell leukemia. As better described in the Methods, once the device is sterilized and filled with growth medium, an adequate number of cells is collected from the culture and centrifuged twice to remove impurities and aggregates. The cells are then manually introduced into the device through a tygon tube connected to the second inlet. Both before and during loading, a constant flow of medium is supplied from the first inlet. As mentioned earlier in section A, this prevents cell sedimentation near the inlet and simultaneously facilitates the entry of cells into the microchannels. Once a sufficient number of cells has been loaded, the tube connected to the second inlet is clamped, and only the flow from the first inlet is maintained.
Loaded chip is placed in a cell incubator and monitored for 4 days taking confocal microscopy images every other minute. 
Figure~\ref{fig:3}a shows a snapshot of a portion of the device. Trapped cells can be easily identified and segmented (see Methods for details on the computational pipeline). The presence of the traps allows for the tracking of cells up to several generations as shown in Figure~\ref{fig:3}b. In particular, once the channels containing cells that have undergone at least two divisions are identified, the image frames can be segmented to track cell growth and proliferation over time. By compiling the information from the segmentation masks across the timelapse, it is possible to generate plots like those shown in Figure~\ref{fig:3}c. These plots represent the growth and division curves for specific branches of the same lineage tree (Figure~\ref{fig:3}d).

Starting from a single progenitor cell, each division creates a new bifurcation in the diagram. By assembling all division events, the complete lineage tree can be reconstructed. Conversely, by selecting a terminal cell and tracing back to the original progenitor, one can isolate the corresponding branch of the tree.

As an example, Figure~\ref{fig:3}c shows the growth curves for the branches leading to the terminal cells highlighted in dark green, purple, and blue. The curves report cell size as a function of time. Beginning with the progenitor cell (brown), at is division only one daughter cell is shown (in red or green), tracked from birth to its own division, and so on. In each graph, the average size at birth (blue dashed line) and at division (red dashed line) are also indicated.

It is worth noting how individual growth curves deviate from these average values. This variation arises from two main factors: first, the tracked cells may not be in a homeostatic phase; and second, the influence of partitioning noise. 

\subsection{Direct measurement of cell proliferation features}

Analyzing the obtained lineages (Figure~\ref{fig:3}), it is possible to extract information about the growth and division processes of single cells. 
As discussed in the previous section, a single timelapse experiment permits the following of hundreds of division events, from which one can readily extract statistical significant information on the growth and division strategy adopted by the analyzed cells. 
Figure~\ref{fig:4}a reports the distribution of the inherited fraction of cell size at division. 
To reconstruct this distribution, it is necessary to measure the size of each daughter cell at birth. In order to define a consistent reference point across all cells, we set the birth size as the area of the daughter cells measured 40 minutes after the cytokinesis of the mother cell. Similarly, the division size of a given cell is defined as its area 60 minutes before undergoing cytokinesis \cite{cadart2018size, zlotek2015optical}.

Once all division events were identified, regardless of when or where they occurred in the device, or whether the daughter cells subsequently divided, we segmented the corresponding frames to determine the size of both daughter cells. The fraction of cell size inherited at birth was then computed as 
$f_{d1} = \frac{\mathrm{Area}_{d1}}{\mathrm{Area}_{d1} + \mathrm{Area}_{d2}}$ and 
$f_{d2} = 1 - f_{d1}$.
To assess the degree of symmetry, the resulting distribution was fitted with the superposition of two Gaussian functions~\cite{Caudo2025}. In the case of asymmetric division, the fit yields two peaks centered at $p$ and $1 - p$; for a symmetric division, a single Gaussian centered at $p=0.5$ is expected. In our dataset, the fit returned a single Gaussian with $\mu =0.5$ and $\sigma=0.05$, consistent with symmetric divisions. Interestingly, the measured asymmetry coefficient is in accordance with the measured partition asymmetry of the cytoplasm of the same kind of cells obtained via flow cytometry~\cite{Peruzzi2021, Miotto2025phyl}. \\

By analyzing all cells for which the complete cell cycle, from birth to division, could be reconstructed, we can infer the growth-control strategy of this cell line.
Cells must coordinate their growth and division to ensure size homeostasis and maintain functionality across generations \cite{taheri2015cell,cadart2018size}. To describe how cells regulate this balance, three main conceptual models have been proposed \cite{facchetti2017controlling,amir2014cell}. In the sizer model, a cell divides once it has reached a critical size threshold, thereby ensuring that excessively small cells grow proportionally more before division, while larger cells divide sooner. The adder model assumes that cells increase their volume by a relatively constant amount between birth and division, regardless of their initial size, a mechanism that naturally stabilizes cell dimensions over successive generations. In contrast, the timer model suggests that division occurs after a defined time interval, such that the extent of growth depends primarily on the duration of the cell cycle rather than on size checkpoints. While these models provide useful conceptual frameworks, experimental evidence indicates that many organisms do not strictly follow a single mechanism but rather display mixed or context-dependent behaviors, suggesting that size control emerges from the interplay of multiple regulatory strategies \cite{nieto2020unification,delarue2017simple}.
Knowing the relationship between the cell size at birth and the change in size from birth to division ($\Delta s$) makes it possible to identify the growth model followed by cells. Specifically, the type of correlation between these two variables distinguishes the models: an adder model corresponds to no correlation (cells add the same amount of volume before division, regardless of their birth size); a sizer model is indicated by a negative correlation (smaller cells must grow more than larger ones to reach the same division size); and a timer model is reflected by a positive correlation (assuming exponential growth, smaller cells grow less than larger ones over the same time span).

We segmented all cells that divided at least twice (a necessary condition to determine both birth and division sizes) and compiled the resulting data for Jurkat cells (Figure~\ref{fig:4}b). All sizes were normalized by the mean birth size. The regression line, also shown in the plot, has a slope of $m = - 0.24$, indicating a sizer-like division strategy, while purely sizer-controlled mechanism would correspond to $m = -1$.
This observation is consistent with previous results obtained from indirect measurements via flow cytometry~\cite{miotto2023determining}.

Note that we are measuring size in terms of cell areas at mid-height, rather than volumes, thus the added size $\Delta S$ is expected to be lower that 1 for $S_b = 1$. In fatc, assuming spherical cells, the relation 
$V_{d} = 2V_{b}$ corresponds to $A_d = 2^{\frac{2}{3}} A_b$, i.e., $\Delta A = 0.587$. Given that cells in the microchannels are not perfectly spherical, our results are in reasonable agreement with theoretical expectations.

Next, we proceeeded to extract the distribution of interdivision times (Figure~\ref{fig:4}c). The distribution peaks at $\Delta h \approx 22 \ \mathrm{h}$, which is in line with the reported doubling time of Jurkat cells in standard culture, considering that growth is known to be affected by the environment~\cite{EnricoBena2021} and microfluidic devices generally prolongs division times by $\sim2~\mathrm{h}$ \cite{nobs2014long,yang2018analysis,tauber2021perform}. Although a secondary peak is observed around $\Delta h \approx 43 \ \mathrm{h}$, the cumulative distribution (inset) shows that $90\%$ of divisions occur within $35 \ \mathrm{h}$.

\subsection{Impact of Channel Geometry and Nutrient Flow on Cell Proliferation Dynamics}
Finally, we investigated the role of channel inclination in cell proliferation to check whether it is possible to retrieve physiological growing conditions by modulating the nutrient flux in the channels. 

To quantify these effects, we reconstructed growth curves for individual cells and grouped the results according to the inclination of the channel from which the data originated. Figure~\ref{fig:5}a shows one representative curve, where cell area was tracked throughout the entire cell cycle. 

Figure~\ref{fig:5}b shows the growth rate distributions stratified by channel inclination. One can see that cells growing in 45° channels exhibit, on average, a higher growth rate than those grown in the other two inclinations. 
This observation supports the hypothesis that the higher flow velocity present in the 45° configuration actively promotes proliferation.
To further check the role of medium flux, we investigated whether and how cells would proliferate in the complete absence of flow. In this scenario, cells inside the device have access only to the medium contained within their own channel and by diffusion to the medium contained in the main channel.

We performed an experiment where Instead of applying a constant flow, we completely stopped medium delivery. Measurements were taken for four days and then analyzed.
 
 Figure~\ref{fig:5}c shows the distributions for cells grown with (green) and without (red) flow. A clear bimodal pattern emerged in the no-flow condition, with two peaks of comparable probability, one centered around 28 h and the other at 46 h. The rightward shift of the first peak relative to the flow condition indicates an overall slowdown in proliferation. The appearance of a pronounced second peak, absent in the flow experiments, can be explained by considering the birth times of the cells.

Figure~\ref{fig:5}d shows division time as a function of birth time (time zero corresponds to the start of data acquisition) for both conditions. Without flow, cells born earlier tend to divide faster than those born later. In contrast, with flow, division times never exceed 40 h, regardless of birth time. Notably, in the flow condition, even second-generation cells (born after 40 h) have division times of around 20 h.

These results suggest that the absence of continuous medium flow plays a key role in limiting proliferation within this type of microfluidic device. Cells that divide shortly after loading, having spent most of their cycle under normal growth conditions, retain a memory of those favorable conditions and divide at rates comparable to those in the flow condition. Conversely, cells that spend their entire cycle without flow take nearly twice as long to grow and divide.

This highlights the essential role of a continuous medium flow within the microfluidic chip to sustain normal cell proliferation.

\section{Conclusions}

In this study, we developed and optimized a microfluidic device designed for the long-term observation of growth and division in suspension T-cells. The device architecture was engineered to allow efficient single-cell trapping and to monitor proliferation across multiple generations by balancing channel dimensions, inclination, and medium flow to ensure favorable conditions both for growth and for automated image analysis.

Our findings show that channel geometry not only affects trapping probability but also impacts cell proliferation dynamics. In particular, channels inclined at 45° exhibited both a higher probability of trapping and higher average growth rates compared with the other configurations. This effect is likely linked to the distinct flow profiles generated along the main channel, which modulate nutrient delivery and waste removal. The analysis of individual growth curves confirmed an exponential growth phase consistent with proliferating T-cells under near-physiological conditions.

The distribution of division times showed a primary peak around 22 h, consistent with the reported doubling time of Jurkat cells in standard culture. 
Analysis of the relationship between birth size and added size until division yielded a regression slope of $m = -0.24$, indicative of a sizer-like growth strategy: smaller cells tend to grow relatively more to reach a more homogeneous division size, although not fully compensatory as in an ideal sizer model (slope -1).

A particularly notable finding concerns the effect of medium flow on proliferation. Experiments performed without continuous flow showed a marked slowdown in growth and division, with division times for cells born in later stages of the experiment nearly doubling compared to those under constant perfusion. This behavior suggests that cells retain the initial conditions when their cycle starts under sufficient nutrient supply, whereas prolonged nutrient deprivation impairs their proliferative capacity. These results underscore that continuous flow is not merely a technical condition of the device but a functional requirement to maintain physiological behavior in long-term experiments.

Overall, these findings demonstrate that our device not only enables the reconstruction of cellular lineage trees and the direct measurement of key parameters such as division time, division symmetry, and growth strategy but also provides a robust experimental platform to investigate how microenvironmental factors, such as channel geometry and medium flow, influence the proliferation of suspension cancer cells. The ability to track single cells over multiple generations, while minimizing mechanical stress and ensuring controlled conditions, represents a significant step toward a quantitative understanding of the mechanisms regulating growth and phenotypic heterogeneity in complex cellular populations.

\section{Materials and Methods}

\subsection*{Device design}
Figure~\ref{fig:1}a1  shows the schematic representation of the proposed microfluidic device.  
This device is characterized by two inlets, one used to inject the buffer into the device and control its flow while the other is used exclusively to load the cells inside the device. The design with two consecutive inlets is intended to minimize cell accumulation around the inlet hole: thus, the buffer flow from the first inlet ensures that all cells enter the main channel.
After the second inlet, there is a region with a series of pillars arranged in staggered rows, the purpose of which is to disperse cells arriving from the inlet and optimize their distribution as they enter the different trapping zones.
Beyond the area with the pillars, the device is divided into three zones, connected to the main channel through three distinct pathways. Each zone consists of a main channel, along whose sides a series of microchannels are arranged in a comb-like pattern.
Depending on the zone, the microchannels are angled differently relative to the main flow channel: from left to right, at 60, 90, and 45 degrees, respectively (Figure~\ref{fig:1}a2). All microchannels have an opening at the end, connecting them to outer channels that run parallel to the main ones and merge back at the end of the microchannel array. The three zones, along with their respective outer channels, converge into a single channel that terminates in an outlet connected to a waste tube.

\subsection*{Microfabrication}
The pattern of microfluidic chip was drawn using KLayout to generate the required photomask. The chip was fabricated by standard soft photolithography techniques.
\subsubsection*{Fabrication of silicon masters}
We briefly explain the steps by which the master mold, i.e. the inverse version of the required microstructures you want to use, is obtained (Figure~\ref{fig:1} b1-b3).
Initially, a 5 cm × 5 cm glass wafer is cleaned to remove dust and any contaminants that could interfere with the fabrication process. A single layer of SU-8 (MicroChem Corp., USA) is then deposited using a spin coater. Specifically, SU-8 3025 is spun at 500 rpm for 10 s, followed by 5000 rpm for 30 s, to achieve the desired thickness of 20 µm (Figure~\ref{fig:1}b1).
The coated wafer undergoes a soft bake at 65°C for 1 min, 95°C for 20 min, and 65°C for 1 min, then is brought to room temperature before being exposed to UV light (UV-KUB2) through the designed photomask (60\% power, 15 s) (Figure~\ref{fig:1}b2). After exposure, a post-exposure bake is performed at 65°C for 1 min, 95°C for 7 min, and 65°C for 1 min.
Finally, the pattern is developed using SU-8 developer (6 min) and then hard-baked following a thermal cycle of 65°C for 1 min, 95°C for 2 min, 150°C for 1.5 h, 95°C for 2 min, and 65°C for 1 min. Thus, the master mold is ready to be used to develop the microfluidic devices (Figure~\ref{fig:1}b3).

\subsubsection*{Fabrication of PDMS devices}
The final PDMS chip is obtained by replica molding from the master mold. To prepare the PDMS mixture, the base elastomer and curing agent (Sylgard 184, Dow Corning) are mixed in a 10:1 ratio and left to degas in a vacuum chamber for at least 30 minutes. The degassed PDMS is then poured over the master mold and placed under vacuum for an additional 10 minutes to remove any remaining air bubbles trapped in the resin (Figure b4).
At this stage, the master mold with PDMS is placed in an oven at 85°C for 30/35 min to cure. Once brought back to room temperature, the cured PDMS layer is carefully peeled off from the master mold (Figure~\ref{fig:1}b5). The edges are trimmed to minimize the risk of delamination, and inlet and outlet holes are punched using a 1.5 mm diameter punch (Figure~\ref{fig:1}b6).
After being cleaned with isopropanol to remove any debris, the PDMS layer is irreversibly bonded to a microscope slide via oxygen plasma surface activation (Plasma Cleaner, Harrick Plasma) (Figures~\ref{fig:1}b7). Finally, the assembled microfluidic chip is left at 65°C for 30 minutes to further enhance adhesion (Figures~\ref{fig:1}b8).

\subsection*{Cell culture} 
E6.1 Jurkat cells were used as a cell model for proliferation study and maintained in RPMI-1640 complete culture medium containing 10\% FBS, penicillin/streptomycin (1/100) plus glutamine (1/100) at 37 C in 5\% CO2. Upon thawing, cells were passaged once before amplification for the experiment. The cells were then harvested, washed and resuspended in the medium to achieve an optimal cell density for the experiment. Once the cells were obtained at an desired cell density, they were drawn into a tygon tube connected to a syringe for loading directly into the cell inlet (Figure~\ref{fig:1}a1). 

\subsection*{Experimental setup}
To remove any air trapped inside the microfluidic channels, the fabricated chip (Figure~\ref{fig:1}c1) is placed in a Petri dish filled with double-filtered water and then moved to a vacuum chamber for at least 30 minutes. Air bubble removal is a crucial step since the presence of microbubbles within the device can compromise the experiment outcome. Finally, the chip is sterilized under UV light for 1 hour.
Two tubes (Tygon Saint-Gobain, ID 0.020 IN, OD 0.060 IN) connected to syringes filled with PBS solution are inserted into the two inlets, buffer and cells, and a manual flow of PBS solution is applied to flush out the water used for channel cleaning. The displaced water is directed into a waste reservoir through a third tube connected to the outlet. The tube inserted into the second inlet is then clamped, while the one in the first inlet is replaced with a tube connected to a 10 mL syringe filled with RPMI-1640 complete medium.
The experimental setup for the device is explained in detail below and illustrated in Figures~\ref{fig:1}c3-c4.
The entire setup is placed inside a top-stage incubator, which regulates temperature and CO2 concentration, and mounted on an Olympus iX73 spinning disk confocal microscope (A) connected to a computer (C). At this point, the syringe containing the medium is mounted onto a syringe pump (E), while the outlet tube is placed inside a waste container (D). Before introducing the cells into the device, a flow of culture medium is applied for a few minutes to remove any remaining PBS. The incubator is also turned on and allowed to reach the growing temperature (37°C).
Cells are manually loaded into the microfluidic device using a syringe (B) connected via a tube to the second inlet.
Cells loading and subsequent trapping into the microchannels are monitored in real time using confocal microscopy at 10x magnification (Figure~\ref{fig:1}c2) for 4 days, taking images every minute.

\subsection*{Image segmentation}
Once the time-lapse acquisition is completed, all data are collected and image segmentation is performed to identify cell divisions.
The first step consists of selecting the channels where a single cell successfully enters and divides. Unsuccessful cases include multiple cells entering the same channel, making it difficult to accurately track the biological processes taking place, no dividing cells (senescent cells), or cells that die after some time.

This initial manual selection is crucial to exclude channels where segmentation would be unreliable, such as when daughter cells overlap inside the channel, when a cell divides into more than two progeny, or when divided cells fuse back together \cite{schmitz2021development}.

The suitable channels are then segmented using a deep-learning-based tool, Cellpose.
Cellpose is a generalist segmentation algorithm trained on a wide variety of biological images. Unlike classical thresholding methods, it uses a convolutional neural network to predict vector flow fields that represent the spatial gradients pointing from each pixel toward the center of the nearest cell. By integrating these flow fields, Cellpose reconstructs precise cell boundaries, even in cases of irregular morphology or variable image quality. This approach allows for highly accurate and robust segmentation across diverse cell types, imaging modalities, and experimental conditions \cite{stringer2021cellpose,pachitariu2022cellpose}.

The segmentation produces masks that uniquely identify each cell in the field of view. From these masks, a broad range of quantitative features can be extracted, including cell perimeter, area, major and minor axis lengths, fluorescence intensity, and any other measurable parameter relevant to the experimental setup.

\subsection*{Image analysis}
All segmentation masks were analyzed using custom Python scripts developed in-house, from which our presented results were obtained. The designed scripts are able to process the masks efficiently and extract a wide range of quantitative features, including cell size, morphology, and time evolution.

The analysis pipeline allowed us to track individual cells across frames, identify division events, and generate statistical distributions describing cell growth and proliferation dynamics. The extracted data were then combined and processed to obtain the plots and quantitative results.

\subsection*{Computational simulations}
In addition to the experimental measurements, we performed computational simulations using COMSOL Multiphysics (version 5.6). The simulations were designed to complement and interpret the experimental data by providing a detailed description of the fluid dynamics and particle behavior inside the microfluidic device.

Specifically, the Laminar Flow module was employed in a shallow-channel configuration to calculate the velocity fields within the device (Fig.~\ref{fig:2}c1). In parallel, the Particle Tracing for Fluid Flow module was used to estimate the probability of cell trapping as a function of the channel inclination (Fig.~\ref{fig:2}c2). In this case, particle motion was modeled under the combined effect of laminar drag forces and Brownian motion, allowing us to capture both deterministic and stochastic aspects of the transport process.

These simulations provided a quantitative framework to interpret the experimental observations and offered insights into how fluid flow and channel geometry affect cell capture efficiency and growth conditions

\section*{Data Availability}

The data that support the findings of this study are available from the corresponding author upon reasonable request.

\section*{Author contributions statement}
M.M., and G.G. conceived research; M.M. supervised the research. G.G. and G.R. contributed additional ideas; S.S. and G.P. performed cell biology experiments;  S.S. and D.C. designed the microfluidic devices; S.S. developed and constructed the microfluidic devices; S.S. and M.M. performed the mother machine experiments; S.S. developed the image-analysis platform.  M.M. and S.S wrote the manuscript; all authors revised the paper.

\section*{Acknowledgments}
This research was partially funded by grants from ERC-2019-Synergy Grant (ASTRA, n. 855923); EIC-2022-PathfinderOpen (ivBM-4PAP, n. 101098989); Project `National Center for Gene Therapy and Drugs based on RNA Technology' (CN00000041) financed by NextGeneration EU PNRR MUR—M4C2—Action 1.4—Call `Potenziamento strutture di ricerca e creazione di campioni nazionali di R\&S' (CUP J33C22001130001). M.P. acknowledges funding from the Italian Ministero dell’Università e della Ricerca under the programme PRIN 2022 ("re-ranking of the final lists"), number 2022KWTEB7, cup B53C24006470006.
G.G. acknowledges funding from H2IOSC Project - Humanities and cultural Heritage Italian Open Science Cloud funded by the European Union – NextGenerationEU – NRRP M4C2 - Project code IR0000029 -CUP B63C22000730005.\\ 
The authors wish to thank the Imaging Facility at Center for Life Nano \& Neuro Science, Istituto Italiano di Tecnologia.

\section*{Competing Interests}
The authors declare no competing interests.

\bibliographystyle{unsrt}
\bibliography{main.bib}

\end{document}